\documentclass[a4paper, 10pt, 5p, preprint, twocolumn]{elsarticle}
\usepackage[utf8]{inputenc}
\usepackage{lipsum}
\usepackage{multicol}
\usepackage{cuted}
\usepackage{hyperref}
\hypersetup{
    colorlinks=true,
    linkcolor=blue,
    filecolor=magenta,      
    urlcolor=cyan,
    pdfpagemode=FullScreen,
}
\usepackage{bm}
\usepackage{placeins}
\usepackage{amsfonts}
\usepackage{multirow}
\usepackage[utf8]{inputenc}
\usepackage{verbatim}
\usepackage{upgreek}
\usepackage{graphicx,color,amsmath,amssymb,amsopn}
\usepackage{hyperref}
\biboptions{sort&compress}

\usepackage{graphicx,animate}
\usepackage[dvipsnames]{xcolor}
\usepackage{soul,xcolor}
\usepackage{xcolor, ulem, color, framed}
\usepackage{float}

\journal{PLB 
}

\begin{document}

\title{Universality of energy-momentum response in kinetic theories}

\author[igfae,bi]{Xiaojian Du} 
\author[bi]{Stephan Ochsenfeld} 
\author[bi]{Sören Schlichting }

\address[igfae]{Instituto Galego de Física de Altas Enerxías (IGFAE), Universidade de Santiago de Compostela, E-15782 Galicia, Spain}
\address[bi]{Fakult\"at f\"ur Physik, Universit\"at Bielefeld, D-33615 Bielefeld, Germany}

\date{\today}

\begin{abstract}
We study the response of the energy-momentum tensor in several kinetic theories, from the simple relaxation time approximation (RTA) to Quantum Chromodynamics (QCD). Irrespective of the differences in microscopic properties, we find a remarkable degree of universality in the response functions from conformal theories. We find that the response to scalar perturbations in kinetic theory can be effectively described by a pair of one hydrodynamic sound mode and one non-hydrodynamic mode. We find that even beyond the range of validity of hydrodynamics, the energy-momentum response in position space can be effectively described by one single mode with non-trivial dispersion relation and residue.
\end{abstract}

\maketitle
\section{Introduction}
Characterizing the macroscopic behavior of quantum many-body systems is of central importance to modern physics, as questions regarding the transport of conserved quantities such as energy, momentum or other conserved charges impact the transport properties of real-world materials as well as the dynamics of the early universe or the space-time evolution of high-energy heavy-ion collisions. Generally, the dynamics of a quantum many-body system can be described from the underlying quantum field theory (QFT), albeit direct calculations in QFT are typically not feasible with present theoretical tools. Hence one commonly resorts to effective descriptions of the real-time dynamics, e.g. in the framework of hydrodynamics or kinetic theory.

By focusing on the dynamics of conserved quantities on long time and large distance scales, it is commonly expected that hydrodynamics emerges as an effective description for any microscopic theory, as explicitly demonstrated for a variety of weakly coupled kinetic theories \cite{Jeon:1995zm,Romatschke:2015gic,Heller:2016rtz,Kurkela:2017xis,Jaiswal:2021uvv} or strongly coupled field theories using holographic methods \cite{Policastro:2002se,Baier:2007ix,Heller:2007qt,Amado:2008ji,Bhattacharyya:2007vjd,Erdmenger:2008rm}. See these reviews \cite{Romatschke:2009im,Florkowski:2017olj} as well. Modern theories of relativistic viscous hydrodynamics include the relaxation of dissipative terms at first~\cite{Kovtun:2019hdm,Bemfica:2019knx,Bemfica:2020zjp} or second order~\cite{Israel:1979wp,Heinz:2005bw,Du:2019obx} in the gradient expansion to preserve causality and stability. However, an effective description in terms of hydrodynamics does not necessarily signal its own breakdown.

Effective kinetic theories of a weakly coupled QFT are valid beyond the long time large distance limit and extend the range of validity of the effective description to far-from-equilibrium systems. By studying the energy-momentum transport in kinetic theory, it is thus possible to obtain insights into the range of applicability of hydrodynamics, as well as insights into the construction of new hydrodynamic effective theories \cite{Hong:2010at,Denicol:2012cn,Ke:2022tqf}.

Notably, questions regarding the applicability of hydrodynamics are also highly relevant from a phenomenological point of view, as hydrodynamic behavior is observed in relativistic heavy-ion collisions (HICs)~\cite{Schenke:2010rr,Song:2010mg,Shen:2011eg,Gale:2013da,Romatschke:2017ejr,Shen:2020mgh}, and even in proton-nucleus collisions where the system size is extremely small and the lifetime of the system is very short. Hence, understanding how the non-equilibrium Quark-Gluon Plasma (QGP) starts to behave like a relativistic viscous fluid \cite{Heller:2015dha,Giacalone:2019ldn,Du:2020zqg,Du:2022bel,Chattopadhyay:2021ive} and to what extent an effective description in terms of relativistic viscous hydrodynamics is applicable in various collision systems~\cite{Ambrus:2022qya,Ambrus:2022koq} is of utmost importance.

In this letter, we will use kinetic theory~\cite{Arnold:2002zm} to analyze the energy-momentum transport in terms of response functions of the energy-momentum tensor in the sound channel. We will consider the most commonly studied models in high-energy nuclear physics, namely scalar $\phi^4$ theory \cite{Mullins:2022fbx}, SU(3) Yang-Mills gauge theory \cite{Xu:2004mz,Xu:2014ega,Kurkela:2015qoa,Keegan:2016cpi,Kurkela:2018wud,Kurkela:2018vqr,Blaizot:2017ucy,Almaalol:2020rnu,Fu:2021jhl,BarreraCabodevila:2022jhi} and quantum chromodynamics (QCD) \cite{Kurkela:2018oqw,Du:2020dvp,Mehtar-Tani:2022zwf} as well as the relaxation time approximation (RTA) \cite{Ambrus:2022koq,Rocha:2022crt,Alalawi:2022pmg,Kamata:2020mka}. By comparing the results for the different microscopic theories, we will demonstrate that the response of the energy-momentum tensor exhibits a universal scaling, which persits even in the presence of large gradients and thus extends well beyond the range of applicability of hydrodynamics.  We will further demonstrate that the universal energy-momentum response in conformal kinetic theories can be modeled rather accurately in terms of one effective sound mode, which dominates on large time and distance scales, and one non-hydrodynamic mode which effectively describes the transient behavior.

\section{Linear response in kinetic theory}
We consider the kinetic theory for scalar $\phi^4$ theory (SCL), SU(3) Yang-Mills gauge theory (YM) and QCD with $N_f=3$ as well as the relaxation time approximation (RTA), such that 
the real-time dynamics of the system is described by the Boltzmann equation
\begin{eqnarray}
p^\mu\partial_\mu f(x,p)&=&C[f(x,p)],
\end{eqnarray}
for the phase-space distribution $f(x,p)$, where $x^{\mu}=(t,\vec{x})$ and $p^{\mu}=(|\vec{p}|,\vec{p})$ denote positions and momenta
of massless on-shell quasi-particles. The collision integrals $C[f(x,p)]$ take the following form for the different microscopic theories
\begin{eqnarray}
C_{\rm RTA}&=&\frac{p^\mu u_\mu}{\tau_R} (f-f_{eq})\;,\\
\nonumber
C_{\rm SCL}&=&\lambda^2 \int d\Omega\left[ f_{1}f_{2}(1+f_3)(1+f_4) - \text{inv.} \right]\;,\\
\nonumber
C_{\rm YM/QCD}&=&C_{\rm YM/QCD}^{1\leftrightarrow2~{\rm inelastic}}+C_{\rm YM/QCD}^{2\leftrightarrow2~{\rm elastic}},
\end{eqnarray}
where the full expressions of collision integrals for Yang-Mills and QCD kinetic theories can be found in~\cite{Du:2020dvp}.\footnote{Note that the convention for the collision integral in this letter differs from~\cite{Du:2020dvp} by a factor $|\vec{p}|$.}

Since the collision integrals $C[f(x,p)]$ vanish under equilibrium conditions $f_{\rm eq}(p)=1/(e^{|\Vec{p}|/T}\pm 1)$, we can consider a spatially homogenous system in thermal equilibrium as our stationary background and study the evolution of linearized perturbations $\delta f(x,p)$ near thermal equilibrium. By decomposing the linearized perturbations into Fourier modes 
\begin{eqnarray}
\delta f(t,\vec{k},p)=\int\frac{d^3x}{(2\pi)^3}e^{-i\vec{k}\cdot \vec{x}}\delta f(x,p).
\end{eqnarray}
with definite wave-number $\vec{k}$, which describes the magnitude of gradients in the system, the evolution is then governed by the linearized Boltzmann equation
\begin{eqnarray}
(p^0\partial_t+i\vec{p}\cdot\vec{k})\delta f(t,\Vec{k},p)&=&\delta C[f_{\rm eq}(p),\delta f(t,\Vec{k},p)].
\label{eq-linear}
\end{eqnarray}
where $\delta C[f_{\rm eq}(p),\delta f(t,\Vec{k},p)]$ denotes the collision integral linearized around the global thermal equilibrium background.

In order to study the energy-momentum response of the system, we will consider an initial perturbation in the scalar/sound channel, which is realized in terms of a linearized temperature perturbation
\begin{eqnarray}
\delta f(t=0,\vec{k},p)=\delta T\frac{\partial f_{\rm eq}(p)}{\partial T}\;.
\end{eqnarray}
By solving Eq.~(\ref{eq-linear}) numerically, as described in detail in~\cite{Du:2020dvp}, we can then compute the time evolution of the distribution $\delta f(t,\Vec{k},p)$, to obtain the resulting perturbation of the energy-momentum tensor as
\begin{eqnarray}
\delta T^{\mu\nu}(\vec{k},t)=\int\frac{d^3p}{(2\pi)^3}\frac{p^{\mu}p^{\nu}}{p^0}\delta f(t,\Vec{k},p),
\end{eqnarray}
and evaluate the response function
\begin{eqnarray}
\tilde{G}^{\mu\nu}_{\alpha\beta}(\vec{k},t)=\frac{\delta T^{\mu\nu}(\vec{k},t)}
{\delta T^{\alpha\beta}(\vec{k},t=0)}\;,
\label{eq-response}
\end{eqnarray}
which simultaneously describes the propagation of initial perturbation and its relaxation towards the equilibrium state.

Since we assume temperature perturbations on a thermal background, the medium is isotropic and the response functions only depend on the magnitude of the wave vector $k=|\vec{k}|$. Without loss of generality, we can orient $\vec{k}$ along the $z$-axis, such that following the nomenclature conventions of \cite{Kurkela:2018vqr}, the nontrivial response functions are the energy response $\tilde{G}_{s}^{s}(k,t)=\tilde{G}_{00}^{00}(k,t)=\frac{\delta T^{00}(k,t)}{\delta T^{00}(k,t=0)}$ and the diagonal terms in tensor response
$\tilde{G}_{s}^{t,\delta}(k,t)=\tilde{G}_{00}^{xx}(k,t)+\tilde{G}_{00}^{yy}(k,t)=\frac{\delta T^{xx}(k,t)+\delta T^{yy}(k,t)}{\delta T^{00}(k,t=0)}$, $\tilde{G}_{s}^{t,k}(k,t)=\tilde{G}_{00}^{zz}(k,t)-\tilde{G}_{00}^{xx}(k,t)-\tilde{G}_{00}^{yy}(k,t)=\frac{\delta T^{zz}(k,t)-\delta T^{xx}(k,t)-\delta T^{yy}(k,t)}{\delta T^{00}(k,t=0)}$
which are related by tracelessness of the energy-momentum tensor $\tilde{G}_{s}^{s}(k,t)=2\tilde{G}_{s}^{t,\delta}(k,t)+\tilde{G}_{s}^{t,k}(k,t)$ in conformal theories.
Another non-trivial term is the vector response $\tilde{G}_{s}^{v}(k,t)=i\tilde{G}_{00}^{0z}(k,t)=i\frac{\delta T^{0z}(k,t)}{\delta T^{0z}(k,t=0)}$ which is related to other response function via energy-momentum conservation $\partial_t\tilde{G}_{s}^{s}(k,t)+k\tilde{G}_{s}^{v}(k,t)=0$, $\partial_t\tilde{G}_{s}^{v}(k,t)-k[\tilde{G}_{s}^{t,\delta}(k,t)+\tilde{G}_{s}^{t,k}(k,t)]=0$.
In order to understand the sound channel in the isotropic medium, the scalar response to the temperature perturbation $\tilde{G}_{s}^{s}(k,t)$ is enough for discussion. We thus put $\tilde{G}_{s}^{t,\delta}(k,t)$,  $\tilde{G}_{s}^{t,k}(k,t)$ and $\tilde{G}_{s}^{v}(k,t)$ in the appendix just for completeness, see Fig.~\ref{fig:ResponseTensorTV}.

\section{Universality of response functions}
Since our aim is to compare macroscopic properties of different microscopic theories it is important to characterize the dynamics in terms of meaningful time and distance scales, that enable a one-to-one comparison.
Since in the long time long wavelength limit hydrodynamics is expected to emerge as the universal low energy effective theory, we can draw guidance from first order relativistic viscous hydrodynamics, where the response function is given by
\begin{align}
\tilde{G}_s^s(k,t)=\tilde{G}_{\rm hydro}^{\rm 1st}(k,t)=\cos(c_skt)e^{-\frac{2}{3}\frac{\eta}{sT}k^2 t},
\end{align}
for conformal theories with $c_s^2=\frac{1}{3}$. Information about the underlying microscopic dynamics is encoded in the shear viscosity $\eta/s$, which depends on the respective coupling strengths, s.t. e.g within RTA one has $\eta/s=\frac{\tau_R T}{5}$, with $\tau_R$ the relaxation time. Similarly, for the other theories, this is shown in Table \ref{tab-etaos}.

\begin{table}[H]
	\center
	\begin{tabular}{|c|c|c|}
		\hline
		Theory & Coupling Parameters & $\eta/s$\\ 
		\hline
		RTA & Relaxation time $\tau_RT$=5& $1.00$\\ 
		\hline
		SCL & Scalar coupling $\lambda\approx$ 83.25 & 1.00 \\ 
		\hline
		YM & 't Hooft coupling $\lambda$=10 & 0.62 \\ 
		\hline
		QCD & 't Hooft coupling $\lambda$=10 & 1.00 \\ 
		\hline
	\end{tabular}
\caption{Summary of $\eta/s$ from different kinetic theories with certain coupling strength or relaxation time.\cite{York:2008rr,Teaney_2014,Moore:2018mma}}
\label{tab-etaos}
\end{table}

\begin{figure}[t!]
    \centering
    \includegraphics[width=0.40\textwidth]{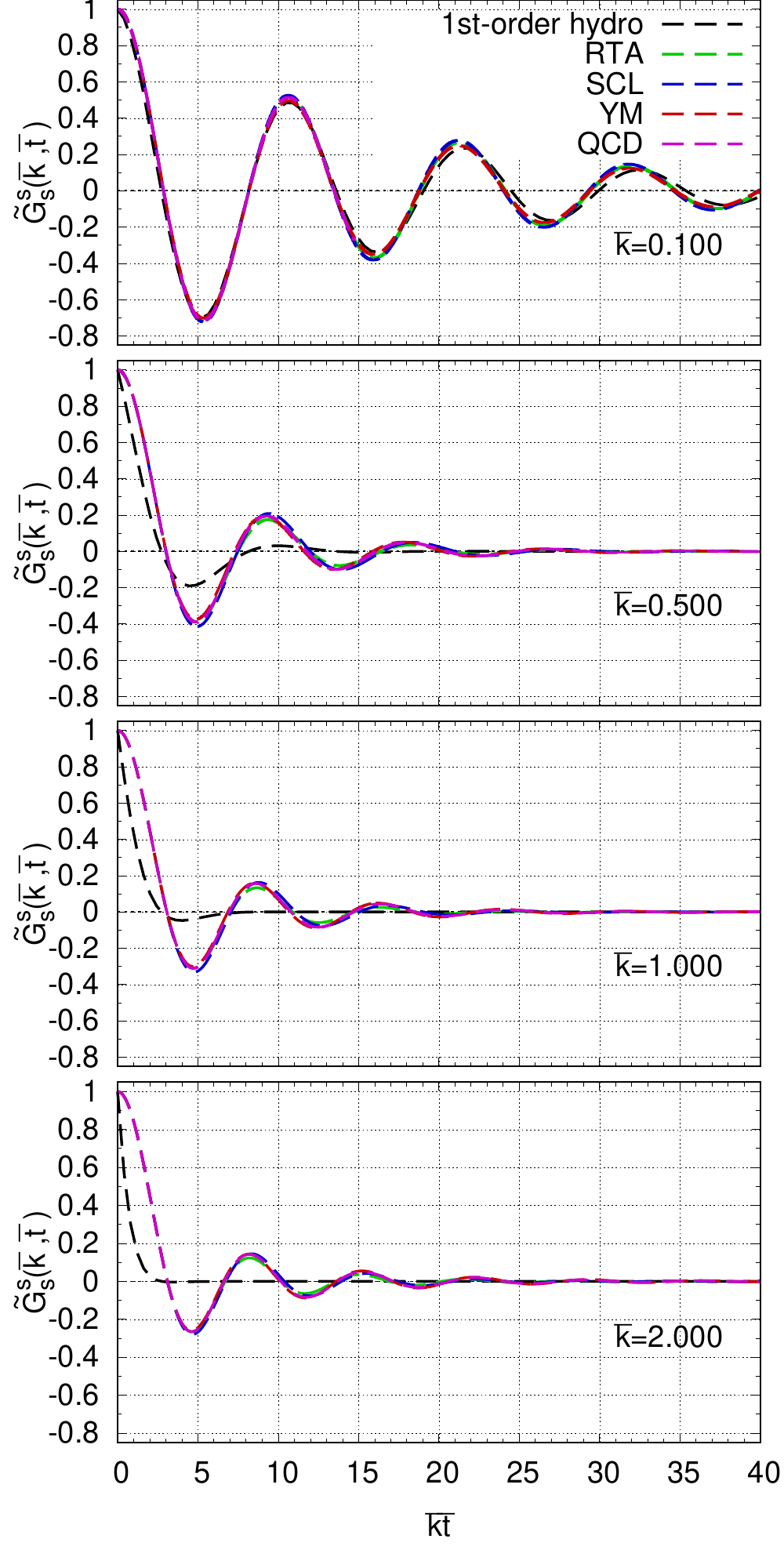}
    \caption{Energy response functions $\tilde{G}_s^s(\bar{k},\bar{t})$ as a function of the propagation phase $\bar{k}\bar{t}$ in different kinetic theories (QCD, YM, SCL, RTA). Different panels show results for different wave-numbers $\bar{k}$. 
    Dashed black curves show the first-order hydrodynamic response function in Eq.~(\ref{eq-hydro-response}).
    \label{fig:Response}
    }
\end{figure}

Now the crucial observation is that the dependence on the coupling strength $\eta/s$ can be eliminated from the hydrodynamic response by defining the scaling variable
\begin{align}
\bar{t}=t\frac{sT}{\eta}\;, \qquad \bar{k}=k\frac{\eta}{sT}\;,
\end{align}
such that the hydrodynamic response assumes a universal form irrespective of the underlying microscopic dynamics
\begin{align}
\tilde{G}_{\rm hydro}^{\rm 1st}(k,t)=\cos(c_s\bar{k}\bar{t})e^{-\frac{2}{3}\bar{k}^2 \bar{t}}\;.
\label{eq-hydro-response}
\end{align}
While this universal scaling emerges naturally in the hydro limit, it is far from obvious that such scaling holds on shorter time ($\Bar{t}\ll1$) and distances ($\Bar{k}\gg1$) scales. \\ \noindent
We analyze this behavior in Fig.~\ref{fig:Response}. Shown are the response functions of the different kinetic theories for various gradients $\Bar{k}$ as colored curves and also the first-order hydrodynamic response in black. For small $\Bar{k}$ we find that the kinetic response functions are described completely by the hydrodynamic Green's function. This also means that we have complete universality between kinetic theories for hydrodynamic scales. Increasing $\Bar{k}$ the hydrodynamic response function strays away from kinetic theories until at large $\Bar{k}$ there is a strong difference between the two. Despite that, we find a striking universality between the kinetic theories to a remarkable degree across all gradient sizes.

\section{Sound and non-hydrodynamic modes}
Now that we have established the universality of the energy-momentum response in kinetic theory, it is interesting to explore to what extent this can be described macroscopically in terms of a simple set of hydrodynamic and non-hydrodynamic excitations of the system. We first note that in first-order viscous hydrodynamics, Green's function in Eq.~(\ref{eq-hydro-response}) can be expressed in terms of a pair of complex conjugated sound modes 
\begin{align}
\label{eq:sound-reponse}
\tilde{G}_s(\bar{k},\bar{t})=\frac{1}{2}\left[Z_s(\Bar{k}) e^{-i[\bar{\omega}_s(\Bar{k}) \Bar{t}+\phi_s(\Bar{k})]}+c.c.\right],
\end{align}
with residue $Z_s(\Bar{k})|_{\rm hydro}^{\rm 1st}=1$, phase $\phi_s(\Bar{k})|_{\rm hydro}^{\rm 1st}=0$ and dispersion relation
\begin{align}
\Bar{\omega}_s(\Bar{k})|_{\rm hydro}^{\rm 1st}=c_s \Bar{k} -i\bar{\Gamma}\Bar{k}^2\;,
\end{align}
where the real part Re$(\Bar{\omega}_s(\Bar{k})/\bar{k})=c_s=1/\sqrt{3}$ yields the propagation speed, while the imaginary part -Im$(\Bar{\omega}_s(\Bar{k}))=\bar{\Gamma}\bar{k}^2=2\bar{k}^2/3$ provides the damping rate, as usual. Second-order Israel-Stuart type hydrodynamic theories, such as BRSSS hydrodynamics \cite{Baier:2007ix} correct the dispersion relation at $\mathcal{O}(\Bar{k}^3)$
\begin{align}
\Bar{\omega}_s(\Bar{k})|_{\rm hydro}^{\rm BRSS}=c_s \bar{k} -i\bar{\Gamma}\bar{k}^2 + \frac{\bar{\Gamma}}{c_s}(c_s^2\bar{\tau}_{\pi}-\frac{\bar{\Gamma}}{2})\bar{k}^3\;,
\label{eq:brsss}
\end{align}
where $\bar{\tau}_{\pi}=\tau_{\pi}\frac{sT}{\eta}=5.1-5.2$ \cite{York:2008rr} and give rise to additional non-hydrodynamic modes with larger decay rates \cite{Israel:1979wp}.

Generally in QFT or in kinetic theory one expects a more complicated singularity structure of the response function in the complex frequency plane \cite{Romatschke:2015gic,Kurkela:2017xis,Moore:2018mma}. Nevertheless, it is conceivable that the essential behavior of the response function can be captured by a finite set of modes. In this spirit, we will describe the kinetic response by one pair of hydrodynamic sound modes and one pair of non-hydrodynamic  modes, i.e.
\begin{eqnarray}
    \Tilde{G}(\bar{k},\bar{t})=\tilde{G}_{s}(\bar{k},\bar{t})+\tilde{G}_{n}(\bar{k},\bar{t})
\end{eqnarray}
where the non-hydrodynamic modes can be expressed in a similar fashion as the hydrodynamic ones
\begin{align}
\label{eq:nonhydro-reponse}
\Tilde{G}_{\rm n}(\bar{k},\bar{t})=\frac{1}{2}\left[Z_n(\Bar{k}) e^{-i[\bar{\omega}_n(\Bar{k}) \Bar{t}+\phi_n(\Bar{k})]}+c.c.\right].
\end{align}
Now in contrast, to Eq.~(\ref{eq:sound-reponse}) where residues, dispersion relations and phase shifts are determined by constitutive relations, we will extract the corresponding coefficients $Z_{s/n},\Bar{\omega}_{s/n}$ and $\phi_{s/n}$ directly from a fit of the real-time response functions in kinetic theory, as detailed below.

Since hydrodynamic and non-hydrodynamic modes always appear in complex conjugated pairs the respective response can be expressed and will be extracted as
\begin{align}
&\Tilde{G}_{s/n}(\bar{k},\bar{t})=\\
& \qquad Z_{s/n}(\Bar{k})\cos[\text{Re}(\bar{\omega}_{s/n}(\Bar{k}))\Bar{t}+\phi_{s/n}(\Bar{k})]e^{\text{Im}(\Bar{\omega}_{s/n}(\Bar{k}))\Bar{t}}. \nonumber 
\end{align}
Since the hydrodynamic sound modes $\Tilde{G}_{s}$ are supposed to describe the late time behavior of the response function, we will extract their features from the late time behavior $\bar{t}>\bar{t}_{s}$ of $\tilde{G}(\bar{k},\bar{t})$. Based on previous studiess~\cite{Giacalone:2019ldn,Du:2020zqg}, which indicate that hydrodynamization typically occurs on time scales $\Bar{t} \approx 4\pi$, it is natural to consider $\Bar{t}_s \sim 4\pi$; in practice,  we will vary $\Bar{t}_s\in [2\pi,6\pi]$ and use the variations of the results to estimate the systematic error. Dispersion relations, residues, and phase shifts are then extracted from the frequency, amplitude, the position of minima, maxima, and zero-crossings of the response function $\Tilde{G}_{s}(\bar{k},\bar{t})$ for $\bar{t}>t_{s}$ as detailed in \ref{sec:extraction}.\\
Non-hydrodynamic modes on the other hand, are supposed to describe the early time behavior and are thus extracted from the behavior of the response function $\Tilde{G}-\Tilde{G}_{s}$ for $\Bar{t}<\Bar{t}_s$. Since for large wave-number $(\bar{k} \gg 1)$ at very early times  $(\bar{t} \ll 1)$, multiple non-hydrodynamic modes can be excited, it also proves necessary to introduce a lower cut-off $\Bar{t}_n$, such that in practice, we limit the extraction to $\bar{t}_{n} < \bar{t} < \bar{t}_{s}$, where $\Bar{t}_n\in [0,0.8\pi]$. Dispersion relations, residues, and phase shifts of non-hydrodynamics modes are then extracted from a least squares fit of $\Tilde{G}-\Tilde{G}_{s}$, where systematic errors are again estimated from variations of the cut-off times $t_{s},t_{n}$.

\begin{figure}[t!]
    \centering
    \includegraphics[width=0.45\textwidth]{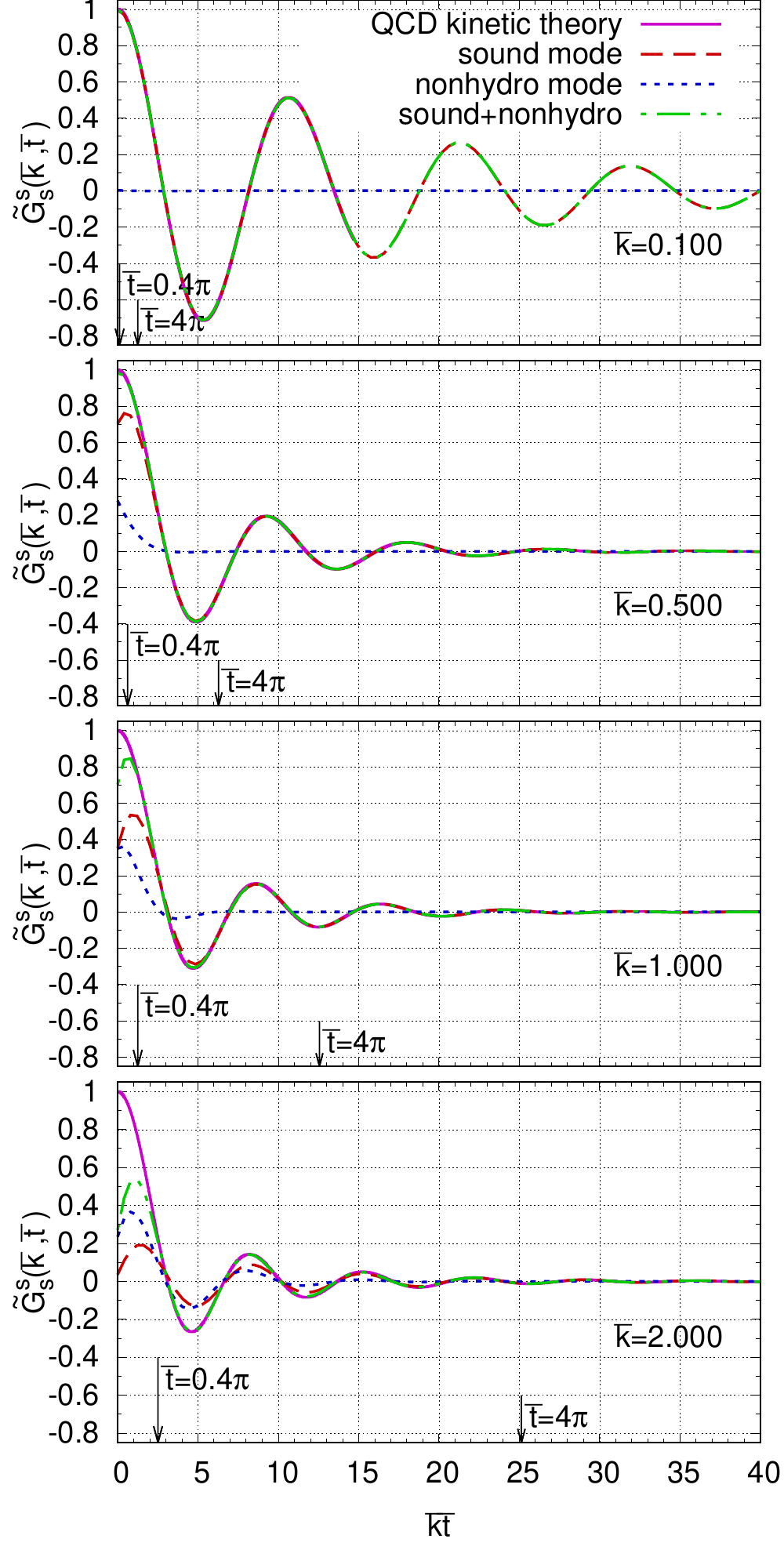}
    \caption{Extraction of effective sound and non-hydrodynamic modes from QCD kinetic theory for different $\bar{k}$ wave modes. Vertical markers indicate extraction windows for sound and non-hydrodynamic modes $\bar{t}=0.4\pi,4\pi$.
    \label{fig:ResponseFit}
    }
\end{figure}

We investigate the quality of this effective macroscopic description in Fig.~\ref{fig:ResponseFit}, where the response function $\Tilde{G}$ obtained from QCD kinetic theory (pink solid), is decomposed into a hydrodynamic (red dashed) and non-hydrodynamic (blue dotted) modes and compared to the sum of the two contributions (green dash-dotted). Starting at small $\bar{k}\lesssim 0.1$, the kinetic response is accurately described by the hydrodynamic sound mode, without any sizeable contribution from the non-hydrodynamic mode. Strikingly, the behavior of the response functions at larger $\bar{k}=0.5,1.0$ is still rather well described by our simple ansatz. However, the non-hydrodynamic modes become more prominent with increasing wave-number $\bar{k}$, such that for $\bar{k}=0.5,1.0$, their contribution at early times is already comparable to that of the hydrodynamic mode. Eventually, for large $\bar{k}=2.0$, the effective description in terms of a single pair of hydrodynamic and non-hydrodynamic modes breaks down, as additional non-hydrodynamic modes should be included to account for the early time behavior of the response function.

\begin{figure}[t!]
    \centering
    \includegraphics[width=0.45\textwidth]{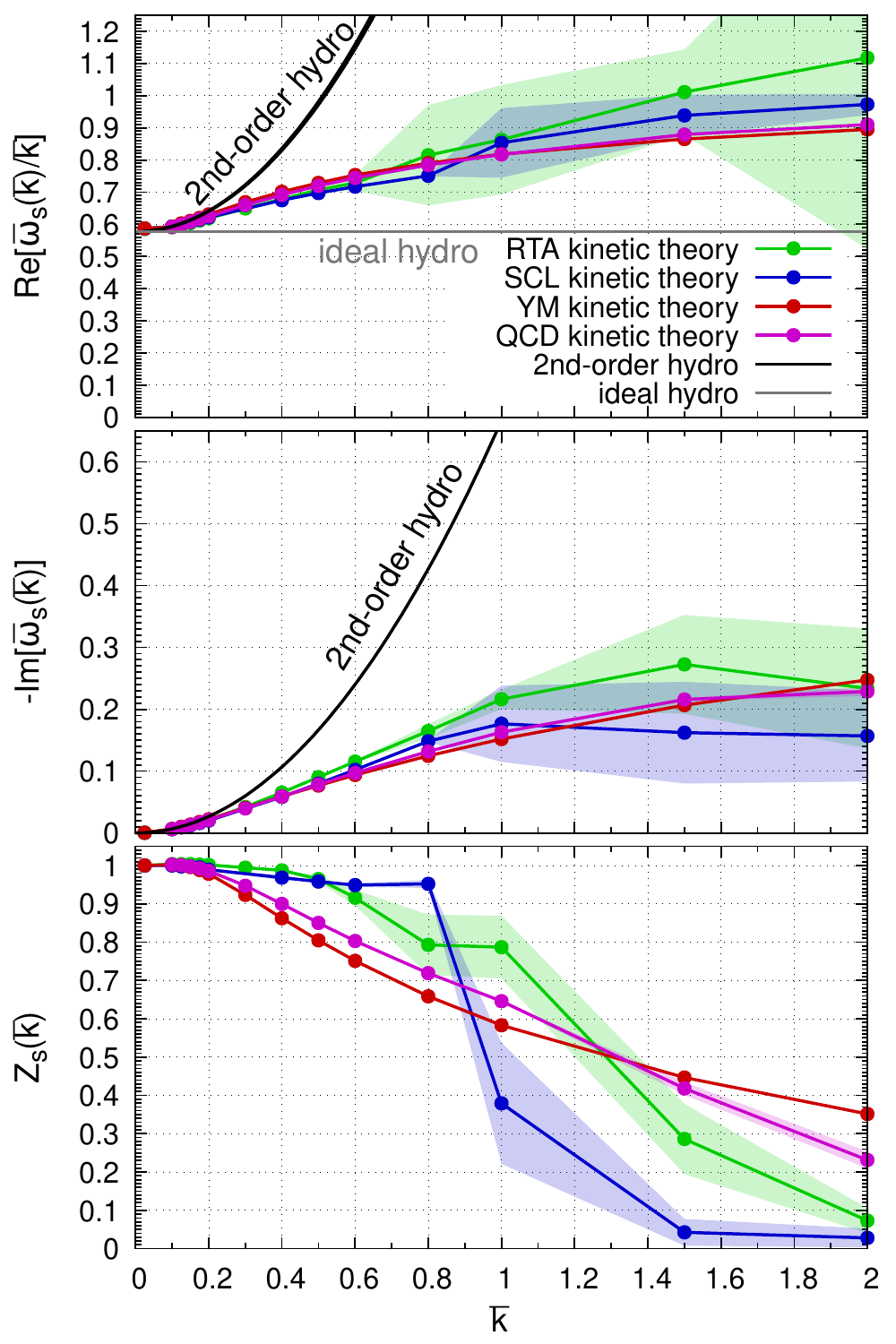}
    \caption{Dispersion relations ${\rm Re}[\omega_s(\bar{k})]/\bar{k}$ (top), $-{\rm Im}[\omega_s(\bar{k})]$ (middle) and residue $Z_s(\bar{k})$ (bottom) for the \textit{effective sound mode} as functions of wave-number $\bar{k}$ for the different kinetic theories (QCD,YM,SCL,RTA). Error bars represent the standard deviation from eleven samples of extraction cuts $\bar{t}_s=2.0\pi,2.4\pi,...,5.6\pi,6.0\pi$. Solid black lines in the top and middle panels indicate the second-order hydrodynamic dispersion relation.
    \label{fig:Dispersion-Sound}
    }
\end{figure}
\begin{figure}[t!]
    \centering
    \includegraphics[width=0.45\textwidth]{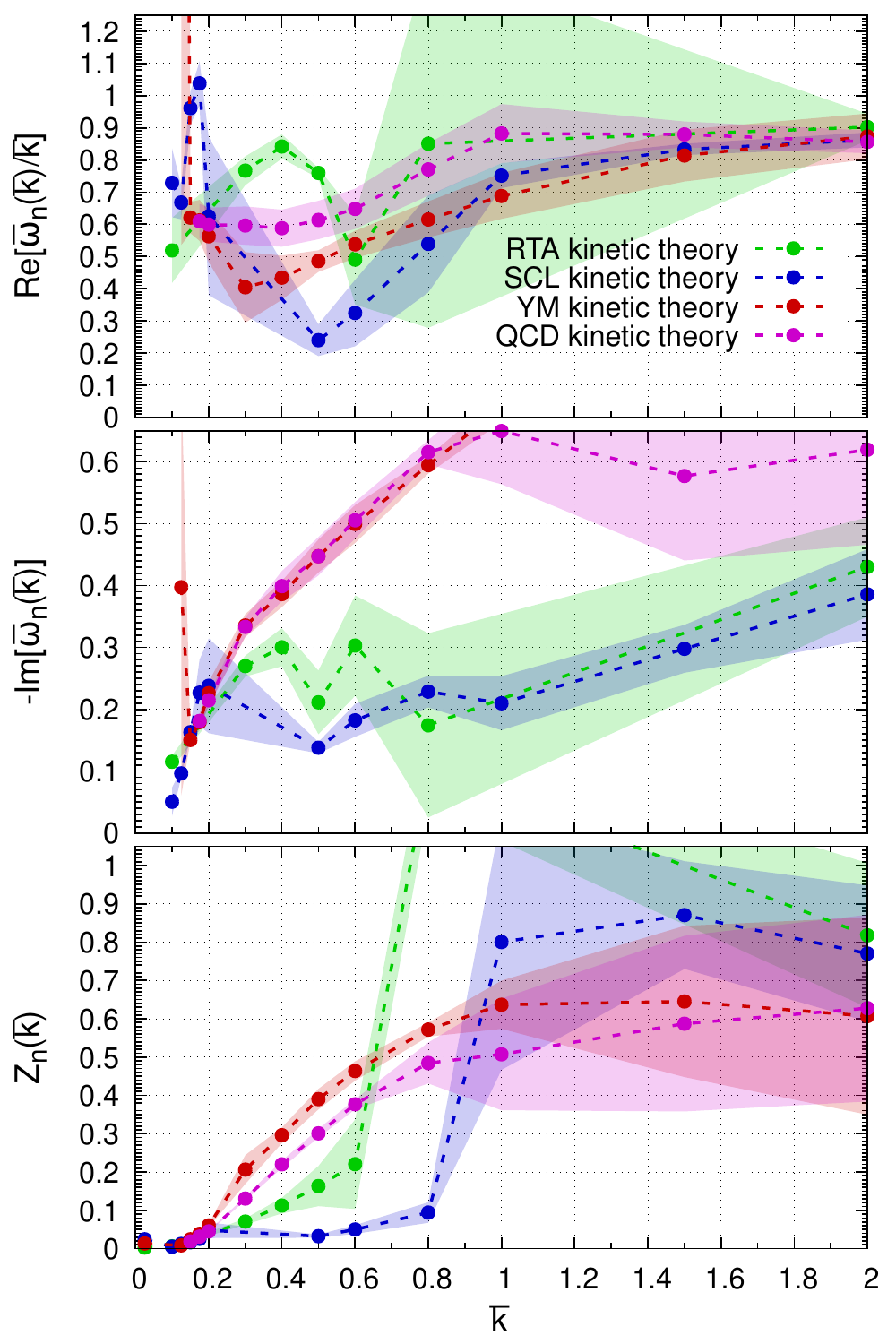}
    \caption{Dispersion relations ${\rm Re}[\omega_n(\bar{k})]/\bar{k}$ (top), $-{\rm Im}[\omega_n(\bar{k})]$ (middle) and residue $Z_n(\bar{k})$ (bottom) for \textit{effective non-hydrodynamic mode} as functions of wave-number $\bar{k}$ for the different kinetic theories (QCD,YM,SCL,RTA). Error bars represent the standard deviation from eleven samples of the extraction cut $\bar{t}_s=2.0\pi,2.4\pi,...,5.6\pi,6.0\pi$ and twenty one samples of the extraction cut $\bar{t}_n=0.00\pi,0.04\pi,...,0.76\pi,0.80\pi$.
    \label{fig:Dispersion-Nonhydro}
    }
\end{figure}

\subsection{Sound modes and hydrodynamics}
The sound modes we just extracted can be directly compared to hydrodynamics and are expected to converge into hydrodynamics at $\bar{k}\ll 1$. 
We present the fitted parameters ${\rm Re}(\bar{\omega}_s(\bar{k}))$, ${\rm Im}(\bar{\omega}_s(\bar{k}))$ and residue $Z_s(\Bar{k})$ of sound modes from kinetic theory as functions of wave-number $\Bar{k}$ in Fig.~\ref{fig:Dispersion-Sound}. We compare the results from RTA (green), scalar $\phi^4$ theory (blue), Yang-Mills kinetic theory (red), and QCD kinetic theory (pink) to ideal hydrodynamics (gray) and second-order BRSSS hydrodynamics (black), see Eq.~(\ref{eq:brsss}). We can directly assess that all kinetic theories converge towards the hydrodynamic solution in the low $\Bar{k}$ limit. The ${\rm Re}(\bar{\omega}_s(\bar{k}))$ values show a strong sense of universality between theories and tend to increase linearly for high $\Bar{k}$. The imaginary part also shows universality among all with larger deviations at large wave-number, where the imaginary part seems to stagnate into a constant value. The residue goes from a clear dominance at low $\Bar{k}$, where hydrodynamics describe the response functions effectively, to small values at large $\Bar{k}$ where the dynamics are taken over by non-hydrodynamic modes.

\subsection{Deciphering the non-hydrodynamic modes}
Now that we discussed the hydrodynamic mode contributions to the kinetic response functions, we proceed to investigate the non-hydrodynamic modes. The dispersion relations $\Bar{\omega}_n(\bar{k})$ and residues $Z_n(\Bar{k})$ are shown in Fig.~\ref{fig:Dispersion-Nonhydro}, where again we compare the results from RTA (green), scalar $\phi^4$ theory (blue), Yang-Mills kinetic theory (red) and QCD kinetic theory (pink). Since the magnitude of the non-hydrodynamic contribution to the response function $Z_n(\Bar{k})$ is very small at small wave-number it is hard to determine the dispersion relation $\Bar{\omega}_n(\bar{k})$ accurately for $\bar{k}\lesssim 0.2$, which gives rise to spurious oscillations in Fig.~\ref{fig:Dispersion-Nonhydro}. Nevertheless, one clearly observes that for all wave-numbers the dispersion relations of the effective non-hydrodynamic modes are characterized by finite real and imaginary parts, with decay rates $-$Im$(\Bar{\omega}_n(\bar{k}))$ exceeding the ones of the hydrodynamic modes. In order to properly interpret this result, it is important to point out that in RTA, it has been established that the kinetic response function exhibits a logarithmic branch cut at Im$(\Bar{\omega}(\Bar{k}))=-0.2$, which in our effective description is approximated by two poles at roughly the same Im$(\Bar{\omega}(\Bar{k}))$. Similarly, also for the other kinetic theories the effective non-hydrodynamic modes should not be regarded as an actual isolated singularity, but rather as an effective description of a more complicated singularity structure in the complex frequency plane~\cite{Moore:2018mma}.

With increasing $\Bar{k}$ the importance $Z_n(\Bar{k})$ of the non-hydrodynamic contribution to the response increases, mirroring the behavior of the residue $Z_s(\Bar{k})$ of the sound mode. When comparing the effective non-hydrodynamic modes for different microscopic theories, one does not observe the same level of universality as for the hydrodynamic modes. Nevertheless, the overall behavior of dispersion relations and residues is still remarkably similar even in the high $\Bar{k}$ regions; in particular QCD and Yang-Mills kinetic theory on the one hand as well as RTA and scalar $\phi^4$ theory on the other hand exhibit essentially the same behavior.
 
\section{Universality of response in position space}
\label{sec:position}
\begin{figure}[t!]
    \centering
    \includegraphics[width=0.45\textwidth]{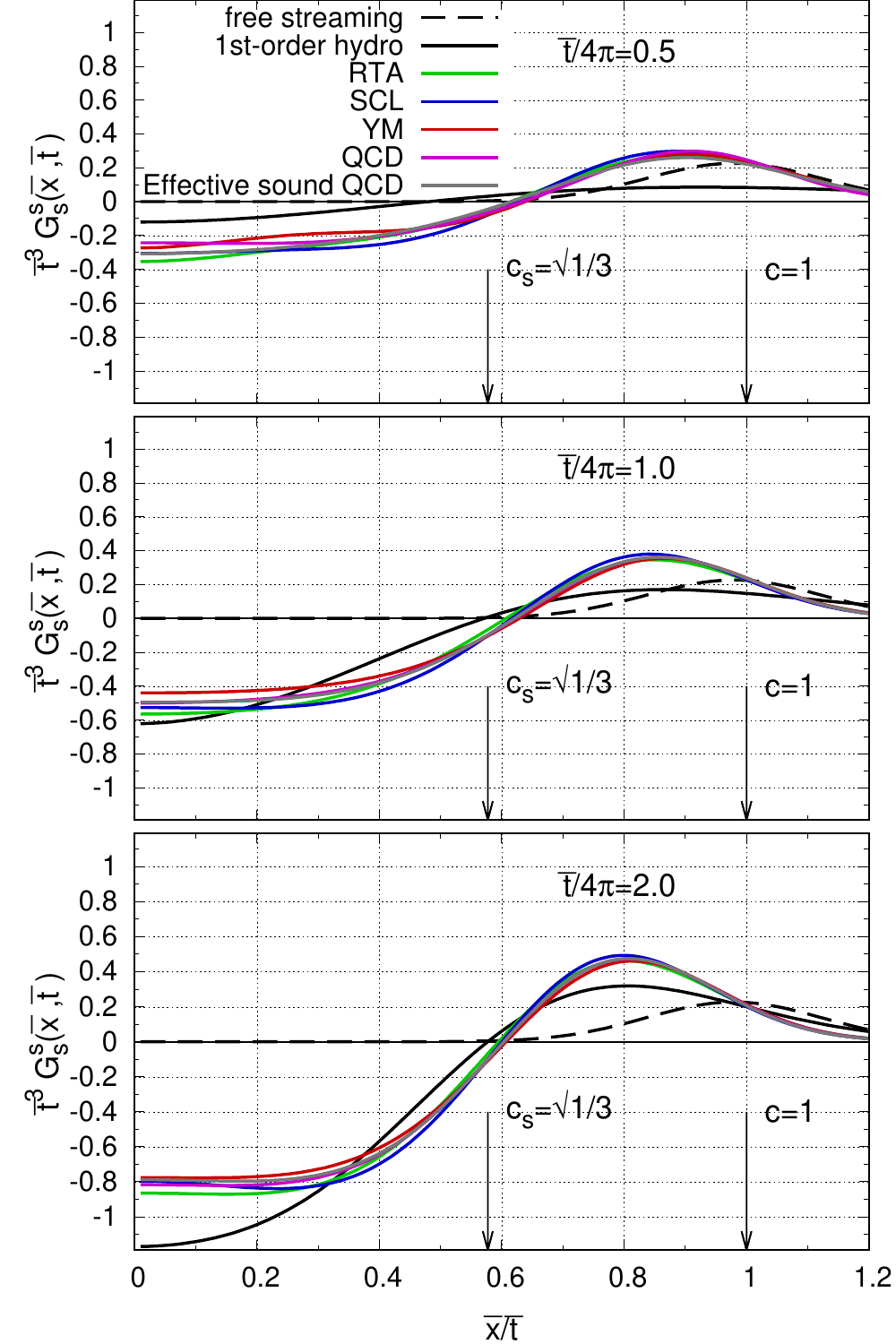}
    \caption{Energy response function $G_{s}^{s}(x,t)$ in position space as a function of the propagation distance $\bar{x}/\bar{t}$ for different kinetic theories (RTA. SCL, YM, QCD). Different panels show the response function at different times $\bar{t}=2\pi,4\pi,8\pi$ in position space for all four kinetic theories (RTA. SCL, YM, QCD). Black solid curves in each panel represent the first-order hydrodynamic response function. Black dashed curves represent the non-interacting free-streaming solution. Gray curves show the reconstruction of the QCD response function from a single effective sound mode, as discussed in Sec.~\ref{sec:position}, which describes the kinetic theory result rather accurately.
    \label{fig:GPosition}
    }
\end{figure}

So far we have analyzed the universality of the energy-momentum response between kinetic theories only in wave-number ($k$) space. Since in practical applications, one is mostly interested in the response in position space, we will now perform a Fourier transform back to position space and discuss the universality of the response function $G_s^s(\Bar{x},\Bar{t})$. We note that in order to stabilize the numerical Fourier transform at large $\Bar{k}$, we employ a Gaussian smearing of the response function, s.t. the position space Green's function are recovered as
\begin{eqnarray}
G_{\alpha\beta}^{\mu\nu}(\vec{x},t)=\int\frac{d^3\bar{k}}{(2\pi)^3}e^{i\vec{k}\cdot\vec{x}}\tilde{G}_{\alpha\beta}^{\mu\nu}(\vec{k},t)e^{-\sigma (\Bar{k}\Bar{t})^2} \ ,
\label{eq:fourierposition}
\end{eqnarray}
where the smearing strength is chosen to be $\sigma=0.01$ in our following analyses. Our results are compactly summarized in Fig.~\ref{fig:GPosition}, where the different panels show the position space Green's functions at three different times $\bar{t}=2\pi,4\pi,8\pi$, with each line corresponding to a different microscopic theory.
We study the various responses as functions of the propagation distance $\Bar{x}/\Bar{t}$, with $\bar{x}=\frac{xT}{\eta/s}$, and evolution time $\bar{t}$, and further contrast the kinetic theory results with the non-interacting free streaming and first order viscous hydrodynamic limit. One observes that at earlier times the kinetic theory curves lie closer to the free streaming peak and subsequently approach the hydrodynamic regime at later times. However, the most striking observation is the level of agreement between the different kinetic theories, where the only visible deviations occur for low $\Bar{x}/\Bar{t}$.

Since we observe this remarkable degree of universality of the response in position $\bar{x}$ space, it is interesting to investigate, to what extent the response function $G_s^s(\Bar{x},\Bar{t})$  can be described in terms of a single effective sound mode. 
In order to construct such an effective sound mode, we fit the dispersion relation ${\rm Re}[\bar{\omega}_s(\bar{k})]$, ${\rm Im}[\bar{\omega}_s(\bar{k})]$, residue $Z_s(\bar{k})$ and phase shift $\phi_s(\bar{k})$, of the sound mode in Fig.~\ref{fig:Dispersion-Sound} with Padé approximants
\begin{align}\label{eq:para1}
{\rm Re}(\Bar{\omega}_s(\Bar{k}))&=c_s\Bar{k}\frac{1+a_2^R\Bar{k}^2+a_4^R\Bar{k}^4}{1+b_2^R\Bar{k}^2+b_4^R\Bar{k}^4} \ , \\ \label{eq:para2}
 {\rm Im}(\Bar{\omega}_s(\Bar{k}))&=-\Gamma \Bar{k}^2\frac{1+a_2^I\Bar{k}^2}{1+b_2^I\Bar{k}^2+b_4^I\Bar{k}^4} \ , \\ \label{eq:para3}
 Z_s(\Bar{k})&=\frac{1}{1+b_2^Z\Bar{k}^2+b_4^Z\Bar{k}^4} \ , \\ \label{eq:para4}
 \phi_s(\Bar{k})&=\frac{a_2^\phi \Bar{k}^2+a_4^\phi \Bar{k}^4}{1+b_2^\phi \Bar{k}^2 +b_4^\phi \Bar{k}^4} \ .
\end{align}
such that for suitable values of the coefficients, the expressions interpolate between hydrodynamic behavior in the limit $\bar{k} \ll 1$ and free-streaming in the limit $\bar{k} \gg 1$. Specifically, one has
\begin{align}
{\rm Re}(\Bar{\omega}_s(\Bar{k}))&\overset{\Bar{k}\ll1}{=}c_s\Bar{k}+(a_2^R-b_2^R)c_s\Bar{k}^3 \ , \\ 
{\rm Re}(\Bar{\omega}_s(\Bar{k})/\Bar{k})&\overset{\Bar{k}\gg1}{=}c_s(a_4^R/b_4^R)=1 \ , \\
{\rm Im}(\Bar{\omega}_s(\Bar{k}))&\overset{\Bar{k}\ll1}{=}-\Gamma \Bar{k}^2 \ .
\end{align}
such that upon fixing the parameter $b_2^R=a_2^R-\bar{\Gamma}(\bar{\tau}_\pi-\frac{\bar{\Gamma}}{2c_s^2})$ with $\bar{\Gamma}=\frac{2}{3}$, $\bar{\tau}_{\pi}=5.1$ the real part of the hydrodynamic dispersion relation is correctly recovered. The hydrodynamic limit for ${\rm Im}(\Bar{\omega}_s(\Bar{k}))$ is already incorporated into the definition of the function, whereas the free-streaming limit for the real part can be enforced by choosing $b_4^R=c_s a_4^R$.

By taking into account these constraints, the Padé approximants are then fitted to properties of the sound mode for QCD Kinetic Theory in Fig.~\ref{fig:Dispersion-Sound}, yielding the following set of parameters
\begin{align}
		a_2^R&=12.20 \ \ a_4^R=8.63 \\
        a_2^I&=9.50 \ \ b_2^I=18.86 \ \ b_4^I=23.11 \\
        b_2^Z&=0.58 \ \ b_4^Z=0.04 \\
        a_2^\phi&=-17.25 \ \ a_4^\phi=-84.97 \ \ b_2^\phi=49.97 \ \ b_4^\phi=49.07
\end{align}
Comparison between the parametrized curves and extracted sound mode values can be seen in \ref{app:para}.
By performing the Fourier transform of the corresponding Green's function 
\begin{align}
G_{\rm s}^{\rm eff}(\Bar{k},\Bar{t})=Z_s(\Bar{k})\cos\left[{\rm Re}(\Bar{\omega}_s(\Bar{k}))\Bar{t}+\phi_s(\Bar{k})\right]e^{{\rm Im}(\Bar{\omega}_s(\Bar{k}))\Bar{t}} \ , 
\end{align}
to position space, as in Eq.~(\ref{eq:fourierposition}), we obtain the effective sound mode for QCD, which is shown as a gray line in Fig.~\ref{fig:GPosition}. We find that the effective sound mode matches the QCD Green's function to a remarkable degree. Discernible differences only appear at early times $\bar{t}$ for small $\Bar{x}/\Bar{t}$, where a hydrodynamic description should expectedly be worse. Nevertheless, it is impressive to see how the entire Green's function can essentially be captured by a single mode, which is smoothly interpolated between the hydrodynamic and free-streaming limit.

\section{Conclusions \& Outlook}
We presented a numerical calculation of the linear response of the energy-momentum tensor in the sound channel for the conformal kinetic theories in the relaxation time approximation (RTA), scalar $\phi^{4}$ theory (SCL), Yang-Mills (YM) and three flavor QCD. 

When described in terms of appropriate scaling variables, the energy-momentum response functions in kinetic theory show a remarkable degree of universality. While such behavior is expected for in the hydrodynamic limit of late times $\bar{t}\gg 1$ and small wave-number $\bar{k} \ll 1$, the universality of the response functions extends even to the large wave-number $\bar{k}\sim 1$ and transient early-time $\bar{t} \sim 1$ region, where hydrodynamics becomes invalid. 

Despite the fact, that Green's functions in kinetic theory are expected to exhibit a rather complicated singularity structure in the complex frequency plane~\cite{Romatschke:2015gic,Kurkela:2017xis,Moore:2018mma}, we find that the energy-momentum response for wave-numbers $\bar{k}\lesssim 2$ and propagation phase $2<\bar{k}\bar{t}<40$ can be effectively modeled in terms of one hydrodynamic and one non-hydrodynamic mode, for which we extracted the dispersion relations and residues. 

Since the effective sound mode provides the dominant contribution at late times $\bar{t}$ and for small wave-numbers $\bar{k}$, its contribution is readily sufficient to describe the energy-momentum response in position space on time scales $\bar{t} \gtrsim 2\pi$. However, to achieve this level of agreement, it is essential to account for the correct form of the dispersion relation, residue, and phase-shift of the effective sound mode, which strongly deviate from the hydrodynamic gradient expansion beyond the small $\bar{k}$ limit.

Since the dispersion relation, residue, and phase-shift of the effective sound mode are essentially universal between the different kinetic theories, it would interesting to investigate to what extent modifications of the hydrodynamic constitutive relations, e.g. by selective inclusion of higher-order gradient corrections, could improve the description for large gradients $\bar{k}$. In this spirit, the results of our microscopic calculation also provide the first guidance for the construction of new effective hydrodynamic descriptions.

\section*{Acknowledgement}
We thank Weiyao Ke, Aleksas Mazeliauskas, Guy D. Moore, Ismail Soudi, Bin Wu, and Yi Yin for their valuable discussions. This work is supported by the Deutsche Forschungsgemeinschaft (DFG) under grant CRC-TR 211 “Strong-interaction matter under extreme conditions” project no. 315477589-TRR 211. XD is also supported by Xunta de Galicia (Centro singular de investigacion de Galicia accreditation 2019-2022), European Union ERDF, the “Maria de Maeztu” Units of Excellence program under project CEX2020-001035-M, the Spanish Research State Agency under project PID2020-119632GB-I00, and European Research Council under project ERC-2018-ADG-835105 YoctoLHC. The authors gratefully acknowledge computing time provided by the Paderborn Center for Parallel Computing (PC2).

\appendix


\section{Nontrivial response functions}

\begin{figure*}[t!]
    \centering
    \includegraphics[width=0.32\textwidth]{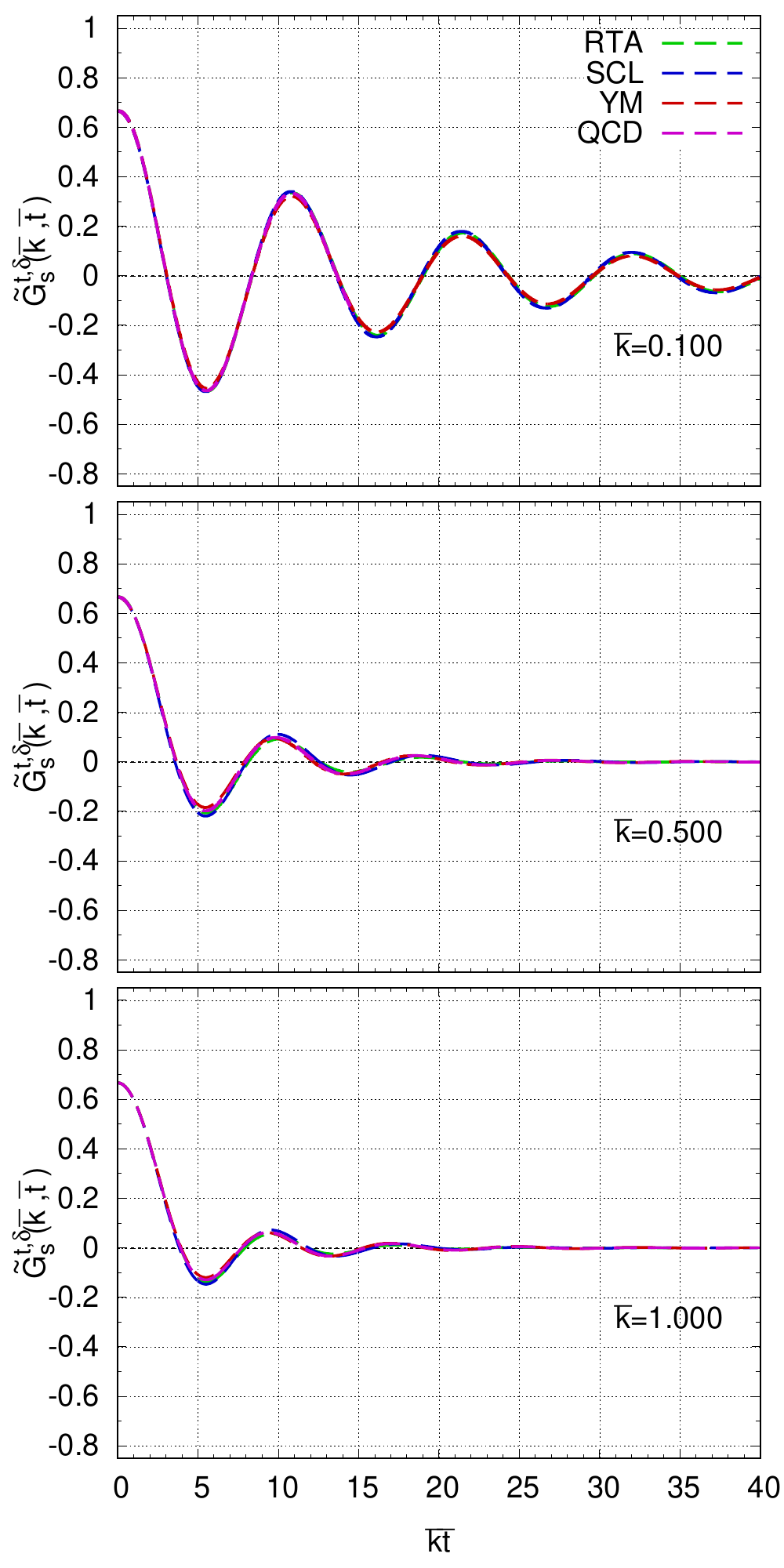}
    \includegraphics[width=0.32\textwidth]{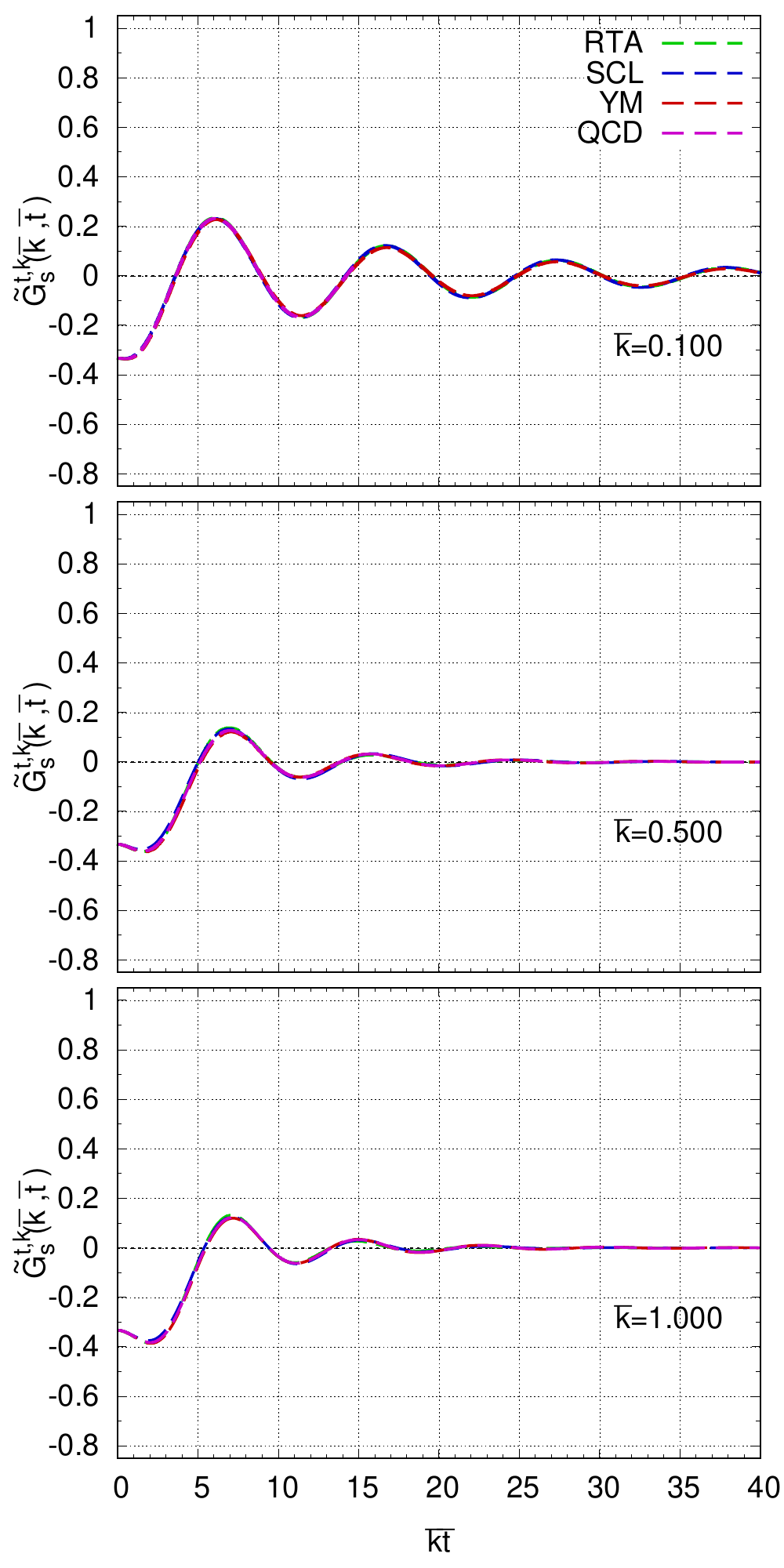}
    \includegraphics[width=0.32\textwidth]{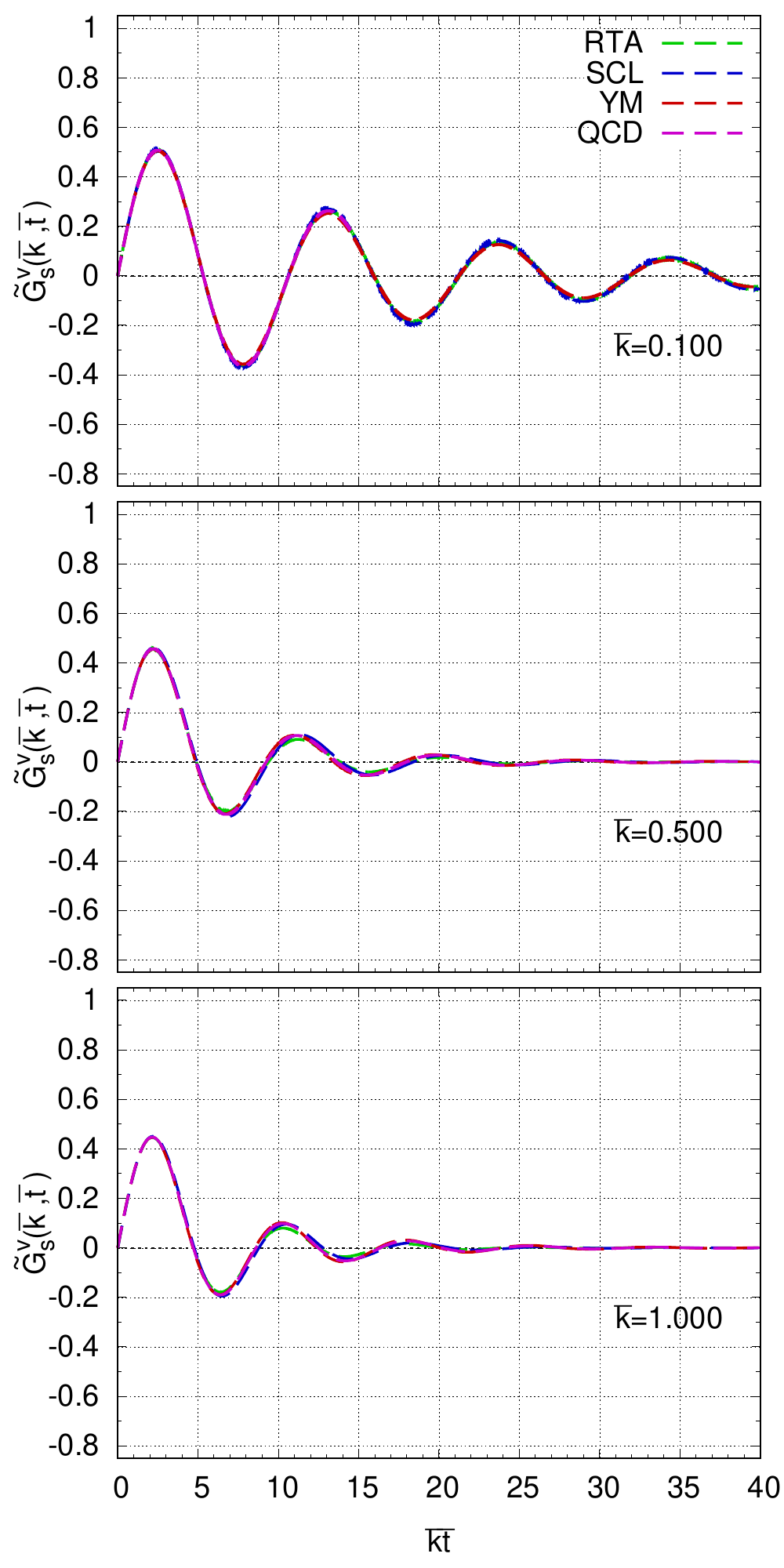}
    \caption{Comparison of the diagonal term in tensor response functions $\tilde{G}_{s}^{t,\delta}(\bar{k},\bar{t})=\tilde{G}_{00}^{xx}(\bar{k},\bar{t})+\tilde{G}_{00}^{yy}(\bar{k},\bar{t})$, $\tilde{G}_{s}^{t,k}(\bar{k},\bar{t})=\tilde{G}_{00}^{zz}(\bar{k},\bar{t})-\tilde{G}_{00}^{xx}(\bar{k},\bar{t})-\tilde{G}_{00}^{yy}(\bar{k},\bar{t})$, and the vector response functions $\tilde{G}_{s}^{v}(\bar{k},\bar{t})=i\tilde{G}_{00}^{0z}(\bar{k},\bar{t})$ between all four kinetic theories (QCD, YM, SCL, RTA) for different $\bar{k}$ wave modes due to temperature perturbation.
    \label{fig:ResponseTensorTV}
    }
\end{figure*}

For the completeness of the sound channel calculation, we present the diagonal terms in the tensor response and the vector response to scalar perturbation.
They have simple relations to each other due to the tracelessness of conformal theories and energy-momentum conservation, which holds in universal scales as well 
\begin{eqnarray}
\tilde{G}_{s}^{s}(\bar{k},\bar{t})-2\tilde{G}_{s}^{t,\delta}(\bar{k},\bar{t})-\tilde{G}_{s}^{t,k}(\bar{k},\bar{t})&=&0\\
\partial_{\bar{t}}\tilde{G}_{s}^{s}(\bar{k},\bar{t})+\bar{k}\tilde{G}_{s}^{v}(\bar{k},\bar{t})&=&0\\
\partial_{\bar{t}}\tilde{G}_{s}^{v}(\bar{k},\bar{t})-\bar{k}\tilde{G}_{s}^{t,\delta}(\bar{k},\bar{t})-\bar{k}\tilde{G}_{s}^{t,k}(\bar{k},\bar{t})&=&0
\end{eqnarray}
The time evolution of these response functions is presented in Fig.~\ref{fig:ResponseTensorTV}.

\section{Extraction algorithm}
\label{sec:extraction}
In the following, we explain the extraction method of the sound mode parameters in detail. \\
The real part of the complex frequency for the sound mode determines the zero crossings of the function, so we can extract it from counting the distance between neighboring zeros in the response function
\begin{eqnarray}
{\rm Re}[\bar{\omega}_s(\bar{k})]=\frac{1}{N}\sum_{j=1}^{N}\frac{\pi}{(\bar{t}^{\rm zero}_{j+1}-\bar{t}^{\rm zero}_{j})}.
\end{eqnarray}
The imaginary part of the complex frequency can be identified with the decay of the response function. Thus, we evaluate the decay rate from neighboring peaks, taking into account their absolute value to ignore the fluctuating sign of the cosine
\begin{eqnarray}
{\rm Im}[\bar{\omega}_s(\bar{k})]=-\frac{1}{N}\sum_{j=1}^{N}\frac{\ln\left( \left|\tilde{G}(\bar{k},\bar{t}^{\rm peak}_{j+1})/\tilde{G}(\bar{k},\bar{t}^{\rm peak}_{j})\right|\right)}{\bar{t}^{\rm peak}_{j+1}-\bar{t}^{\rm peak}_{j}}.
\end{eqnarray}
The phase can also be extracted with help of peaks, as in the phase determines how much the peak is shifted from $\pi$ or $-\pi$ (depending on sign). We take the peak time $\Bar{t}_j^{\rm peak}$ and have to keep in mind that the decay also shifts the peaks slightly by the value $\arctan\left({\rm Im}(\bar{\omega}_s(\bar{k}))/{\rm Re}(\bar{\omega}_s(\bar{k}))\right)$. Thus each peak gives a phase of
\begin{eqnarray}
\phi_s^{j}(\bar{k})=-{\rm Re}[\bar{\omega}_s(\bar{k})] \bar{t}^{\rm peak}_{j}-\arctan\left(\frac{{\rm Im}[\bar{\omega}_s(\bar{k})]}{{\rm Re}[\bar{\omega}_s(\bar{k})]}\right),
\end{eqnarray}
which needs to be averaged later on in the unit circle. \\
The residue can be reconstructed from the peak heights and the decay rate. Since the peak is shifted we also have to account that the cosine is not always 1 or -1 at the peak position so we have to divide by an extra factor. With this in mind, the residue can be calculated by 
\begin{eqnarray}
Z_s(\bar{k})=\frac{1}{N}\sum_{j=1}^{N}\frac{\tilde{G}(\bar{k},\bar{t}^{\rm peak}_{j})e^{{\rm Im}[\bar{\omega}_s(\bar{k})]\bar{t}^{\rm peak}_{j}}}{\cos({\rm Re}[\bar{\omega}_s(\bar{k})]\bar{t}^{\rm peak}_{j}+\phi_s^{j}(\bar{k}))}.
\end{eqnarray}

\section{Parametrization of Sound Mode}\label{app:para}
\begin{figure}[t!]
    \centering
    \includegraphics[width=0.45\textwidth]{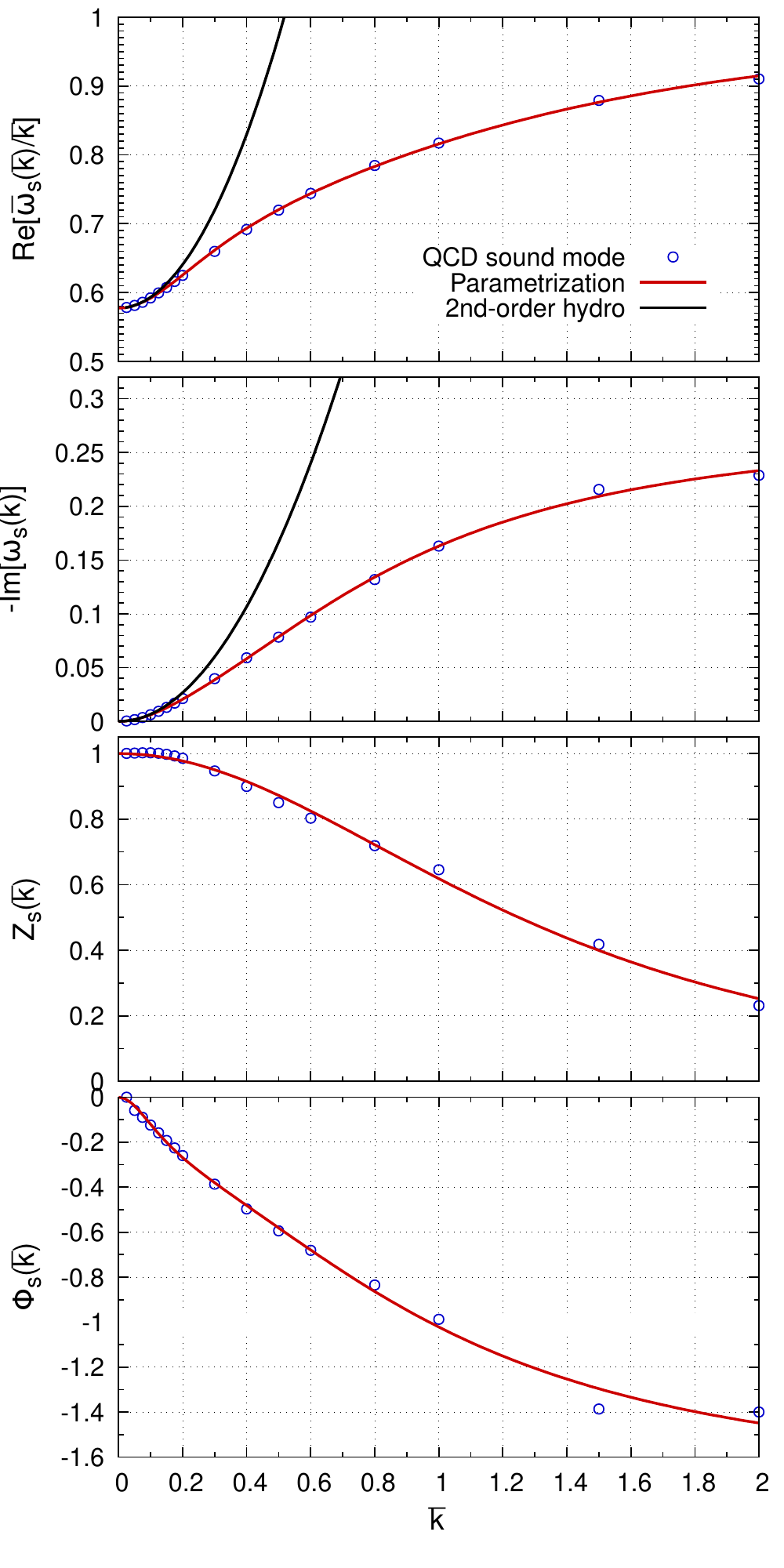} 
    \caption{Fitted curves to sound mode properties. Blue circles are extracted sound mode properties and red curves are fitted to this data.}
    \label{fig:ParaCurves}
\end{figure}

The parametrized curves of the sound mode properties can be seen in Fig.~\ref{fig:ParaCurves}. Each blue circle displays a fitted parameter for certain $\Bar{k}$ and the red curves are the corresponding parametrized curves from Eq.~(\ref{eq:para1})-(\ref{eq:para4}). For reference, there are also black curves that show the second-order hydrodynamic mode. The parametrized curves fit very well with the data overall and clearly reproduce the low $\Bar{k}$ behavior, which they have been built to do.

\bibliographystyle{h-elsevier}

\bibliography{ref.bib}

\begin{thebibliography}{10}

\bibitem{Jeon:1995zm}
S. Jeon and L.G. Yaffe,
\newblock Phys. Rev. D 53 (1996) 5799, hep-ph/9512263.

\bibitem{Romatschke:2015gic}
P. Romatschke,
\newblock Eur. Phys. J. C 76 (2016) 352, 1512.02641.

\bibitem{Heller:2016rtz}
M.P. Heller et~al.,
\newblock Phys. Rev. D 97 (2018) 091503, 1609.04803.

\bibitem{Kurkela:2017xis}
A. Kurkela and U.A. Wiedemann,
\newblock Eur. Phys. J. C 79 (2019) 776, 1712.04376.

\bibitem{Jaiswal:2021uvv}
S. Jaiswal et~al.,
\newblock Phys. Rev. C 105 (2022) 024911, 2107.10248.

\bibitem{Policastro:2002se}
G. Policastro, D.T. Son and A.O. Starinets,
\newblock JHEP 09 (2002) 043, hep-th/0205052.

\bibitem{Baier:2007ix}
R. Baier et~al.,
\newblock JHEP 04 (2008) 100, 0712.2451.

\bibitem{Heller:2007qt}
M.P. Heller and R.A. Janik,
\newblock Phys. Rev. D 76 (2007) 025027, hep-th/0703243.

\bibitem{Amado:2008ji}
I. Amado et~al.,
\newblock JHEP 07 (2008) 133, 0805.2570.

\bibitem{Bhattacharyya:2007vjd}
S. Bhattacharyya et~al.,
\newblock JHEP 02 (2008) 045, 0712.2456.

\bibitem{Erdmenger:2008rm}
J. Erdmenger et~al.,
\newblock JHEP 01 (2009) 055, 0809.2488.

\bibitem{Romatschke:2009im}
P. Romatschke,
\newblock Int. J. Mod. Phys. E 19 (2010) 1, 0902.3663.

\bibitem{Florkowski:2017olj}
W. Florkowski, M.P. Heller and M. Spalinski,
\newblock Rept. Prog. Phys. 81 (2018) 046001, 1707.02282.

\bibitem{Kovtun:2019hdm}
P. Kovtun,
\newblock JHEP 10 (2019) 034, 1907.08191.

\bibitem{Bemfica:2019knx}
F.S. Bemfica et~al.,
\newblock Phys. Rev. D 100 (2019) 104020, 1907.12695,
\newblock [Erratum: Phys.Rev.D 105, 069902 (2022)].

\bibitem{Bemfica:2020zjp}
F.S. Bemfica, M.M. Disconzi and J. Noronha,
\newblock Phys. Rev. X 12 (2022) 021044, 2009.11388.

\bibitem{Israel:1979wp}
W. Israel and J.M. Stewart,
\newblock Annals Phys. 118 (1979) 341.

\bibitem{Heinz:2005bw}
U.W. Heinz, H. Song and A.K. Chaudhuri,
\newblock Phys. Rev. C 73 (2006) 034904, nucl-th/0510014.

\bibitem{Du:2019obx}
L. Du and U. Heinz,
\newblock Comput. Phys. Commun. 251 (2020) 107090, 1906.11181.

\bibitem{Hong:2010at}
J. Hong and D. Teaney,
\newblock Phys. Rev. C 82 (2010) 044908, 1003.0699.

\bibitem{Denicol:2012cn}
G.S. Denicol et~al.,
\newblock Phys. Rev. D 85 (2012) 114047, 1202.4551,
\newblock [Erratum: Phys.Rev.D 91, 039902 (2015)].

\bibitem{Ke:2022tqf}
W. Ke and Y. Yin,
\newblock Phys. Rev. Lett. 130 (2023) 212303, 2208.01046.

\bibitem{Schenke:2010rr}
B. Schenke, S. Jeon and C. Gale,
\newblock Phys. Rev. Lett. 106 (2011) 042301, 1009.3244.

\bibitem{Song:2010mg}
H. Song et~al.,
\newblock Phys. Rev. Lett. 106 (2011) 192301, 1011.2783,
\newblock [Erratum: Phys.Rev.Lett. 109, 139904 (2012)].

\bibitem{Shen:2011eg}
C. Shen et~al.,
\newblock Phys. Rev. C 84 (2011) 044903, 1105.3226.

\bibitem{Gale:2013da}
C. Gale, S. Jeon and B. Schenke,
\newblock Int. J. Mod. Phys. A 28 (2013) 1340011, 1301.5893.

\bibitem{Romatschke:2017ejr}
P. Romatschke and U. Romatschke,
\newblock {Relativistic Fluid Dynamics In and Out of Equilibrium}Cambridge
  Monographs on Mathematical Physics (Cambridge University Press, 2019),
  1712.05815.

\bibitem{Shen:2020mgh}
C. Shen and L. Yan,
\newblock Nucl. Sci. Tech. 31 (2020) 122, 2010.12377.

\bibitem{Heller:2015dha}
M.P. Heller and M. Spalinski,
\newblock Phys. Rev. Lett. 115 (2015) 072501, 1503.07514.

\bibitem{Giacalone:2019ldn}
G. Giacalone, A. Mazeliauskas and S. Schlichting,
\newblock Phys. Rev. Lett. 123 (2019) 262301, 1908.02866.

\bibitem{Du:2020zqg}
X. Du and S. Schlichting,
\newblock Phys. Rev. Lett. 127 (2021) 122301, 2012.09068.

\bibitem{Du:2022bel}
X. Du et~al.,
\newblock Phys. Rev. D 106 (2022) 014016, 2203.16549.

\bibitem{Chattopadhyay:2021ive}
C. Chattopadhyay et~al.,
\newblock Phys. Lett. B 824 (2022) 136820, 2107.05500.

\bibitem{Ambrus:2022qya}
V.E. Ambrus, S. Schlichting and C. Werthmann,
\newblock Phys. Rev. Lett. 130 (2023) 152301, 2211.14356.

\bibitem{Ambrus:2022koq}
V.E. Ambrus, S. Schlichting and C. Werthmann,
\newblock Phys. Rev. D 107 (2023) 094013, 2211.14379.

\bibitem{Arnold:2002zm}
P.B. Arnold, G.D. Moore and L.G. Yaffe,
\newblock JHEP 01 (2003) 030, hep-ph/0209353.

\bibitem{Mullins:2022fbx}
N. Mullins, G.S. Denicol and J. Noronha,
\newblock Phys. Rev. D 106 (2022) 056024, 2207.07786.

\bibitem{Xu:2004mz}
Z. Xu and C. Greiner,
\newblock Phys. Rev. C 71 (2005) 064901, hep-ph/0406278.

\bibitem{Xu:2014ega}
Z. Xu et~al.,
\newblock Phys. Rev. Lett. 114 (2015) 182301, 1410.5616.

\bibitem{Kurkela:2015qoa}
A. Kurkela and Y. Zhu,
\newblock Phys. Rev. Lett. 115 (2015) 182301, 1506.06647.

\bibitem{Keegan:2016cpi}
L. Keegan et~al.,
\newblock JHEP 08 (2016) 171, 1605.04287.

\bibitem{Kurkela:2018wud}
A. Kurkela et~al.,
\newblock Phys. Rev. Lett. 122 (2019) 122302, 1805.01604.

\bibitem{Kurkela:2018vqr}
A. Kurkela et~al.,
\newblock Phys. Rev. C 99 (2019) 034910, 1805.00961.

\bibitem{Blaizot:2017ucy}
J.P. Blaizot and L. Yan,
\newblock Phys. Lett. B 780 (2018) 283, 1712.03856.

\bibitem{Almaalol:2020rnu}
D. Almaalol, A. Kurkela and M. Strickland,
\newblock Phys. Rev. Lett. 125 (2020) 122302, 2004.05195.

\bibitem{Fu:2021jhl}
Y. Fu et~al.,
\newblock Phys. Rev. D 105 (2022) 054031, 2110.01540.

\bibitem{BarreraCabodevila:2022jhi}
S. Barrera~Cabodevila, C.A. Salgado and B. Wu,
\newblock Phys. Lett. B 834 (2022) 137491, 2206.12376.

\bibitem{Kurkela:2018oqw}
A. Kurkela and A. Mazeliauskas,
\newblock Phys. Rev. D 99 (2019) 054018, 1811.03068.

\bibitem{Du:2020dvp}
X. Du and S. Schlichting,
\newblock Phys. Rev. D 104 (2021) 054011, 2012.09079.

\bibitem{Mehtar-Tani:2022zwf}
Y. Mehtar-Tani, S. Schlichting and I. Soudi,
\newblock (2022), 2209.10569.

\bibitem{Rocha:2022crt}
G.S. Rocha et~al.,
\newblock Acta Phys. Polon. Supp. 16 (2023) 29, 2207.11286.

\bibitem{Alalawi:2022pmg}
H. Alalawi and M. Strickland,
\newblock JHEP 12 (2022) 143, 2210.00658.

\bibitem{Kamata:2020mka}
S. Kamata et~al.,
\newblock Phys. Rev. D 102 (2020) 056003, 2004.06751.

\bibitem{York:2008rr}
M.A. York and G.D. Moore,
\newblock Phys. Rev. D 79 (2009) 054011, 0811.0729.

\bibitem{Teaney_2014}
D. Teaney and L. Yan,
\newblock Physical Review C 89 (2014) 014901, 1304.3753.

\bibitem{Moore:2018mma}
G.D. Moore,
\newblock JHEP 05 (2018) 084, 1803.00736.

\end{thebibliography}
\end{document}